# Continuous and Coordinated Efforts of Structure Wakefield Acceleration (SWFA) Development for an Energy Frontier Machine

Chunguang Jing,[i,ii] John Power,[i] Jiahang Shao,[i] Gwanghui Ha,[i] Philippe Piot,[i,iii] , Xueying Lu,[i,iii] Alexander Zholents,[i] Alexei Kanareykin,[ii] Sergey Kuzikov,[ii] James B. Rosenzweig[iv], Gerard Andonian [iv], Evgenya Ivanovna Simakov,[v] Janardan Upadhyay,[v], Chuanxiang Tang,[vi] Richard J Temkin[vii], Emilio Alessandro Nanni [viii], John Lewellen,[viii]

[i] Argonne National Laboratory, Lemont, IL.
[ii] Euclid Techlabs, LLC, Bolingbrook, IL
[iii] Northern Illinois University, Dekalb, IL.
[iv] University of California, Los Angeles
[v] Los Alamos National Laboratory, Los Alamos, NM.
[vi] Tsinghua University, China.
[vii] Massachusetts Institute of Technology, Cambridge, MA.
[viii] SLAC National Accelerator Laboratory, CA

## Table of Contents

# Executive Summary

Structure wakefield acceleration (SWFA) is well suited for the linear collider (LC) application due to its natural ability to accelerate positrons and preserve emittance. Under the SWFA roadmap, which was developed in response to Snowmass 2013 recommendations, four principal technologies—drive beam, main beam, wakefield structure, and LC facility design—have been investigated. The two SWFA schemes under development are the collinear wakefield accelerator (CWA), in which the drive and main beam follow the same path through a structure, and the two-beam accelerator (TBA), where the drive and main beam pass through different structures.

Progress has been made in all principal areas with an accelerated pace in the past decade. Highlights in the TBA research include: 1) increased the acceleration gradient from 10MV/m to 300MV/m; 2) increased wakefield power generation from 50MW to 500MW; 3) demonstrated staged acceleration. Highlights in CWA include: 1) achieved 300MeV/m beam acceleration in a THz structure; 2) transformer ratio reached a new record of 5 with improved longitudinal beam shaping technologies; 3) the theory and technology needed to achieve sustainable THz acceleration has been well-established.

To further advance the SWFA technology in the next decade, continuous and coordinated efforts must be carried out in a more synchronized way. We envision focusing on the following research areas:

- **Implement a fully featured demonstrator in dedicated test facilities**: The key R&D task to realize an SWFA-based facility design is the integration of the sub-components into a self-consistent parameter set needed for the application: a future LC or other non-HEP application (e.g., a compact X-ray source). The design, construction and test of a laboratory-scale SWFA module is needed to demonstrate that SWFA components can be successfully integrated into a working accelerator. This module can then be scaled to higher energies to provide a path towards a future energy frontier machine
- **Synthesize beam generation and manipulation technologies to tailor to a variety of needs for the SWFA**: Drive-electron beam technology is used as the power source for both schemes, although the drive beam format for the two schemes is very different. In the CWA scheme, a single e- drive bunch with a carefully shaped longitudinal profile is used to accelerate a single e- or e+ main bunch, while the TBA scheme requires the generation of a high-current electron drive-bunch train. The main-beam technology ranges from the conventional e+e- damping ring technology to new efforts that use a shaped main beam to significantly increase efficiency.
- **Mature full-featured wakefield structures for sustainable high gradient beam acceleration**: The structure development aims to design reliable wakefield structures capable of high-gradient and high drive-to-main efficiency that accommodates higher order mode damping. Such full-featured wakefield structures can be directly incorporated in a future LC.
- **Perform an integrated design study of the SWFA linear collider scheme**: The SWFA *Science & Technology* has made strong progress since Snowmass 2013 and the 2016 AAC Roadmap but the SWFA strawman design has not been updated during this time. Limited research budgets necessitated focusing on beam and structures but a thorough integrated design study (IDS) is needed to mature the design beyond the strawmand phase. The IDS would consider issues from start to end including: e+ and e- sources, acceleration, beam delivery system, final focus and interaction region.

- **Update the SWFA Roadmap.** Based on the recent rapid progress with SWFA S&T, the SWFA roadmap towards a future linear collider that was originally published in 2016, needs to be revisited and updated in order to best fit to needs of the HEP community in the upcoming decade.

# 1. Introduction

Despite the urgent need for a TeV-class linear collider in high-energy physics, a clear path to practical and affordable accelerator technologies has yet to be realized. In 2016, the United States Department of Energy (DoE) published the "Advanced Accelerator Development Strategy Report" (Advanced Accelerator Concepts Research Roadmap Workshop Report, 2016), in which the conceptual roadmap for the Dielectric Wakefield Accelerator (DWA) was developed in the framework of a decadal timescale. In the 2018 DoE Comparative Review, the concept of the DWA was formally expanded to **SWFA** (structure wakefield accelerators), in order to more broadly and inclusively cover this field of research, as well as to more directly compare with the other two advanced accelerator concepts, PWFA (plasma wakefield accelerators) and LPA (laser plasma accelerators).

The definition of SWFA can be generalized to an acceleration method that uses a particle beam (i.e., the drive beam) as a pump source and solid structures as the medium to enable an energy transfer from the pump source to a to-be-accelerated beam (i.e., the main beam). This definition singles out the two critical elements, the structure and the beam. The structure is used to confine an established electromagnetic wave, which operates in the accelerating mode with a frequency ranging from GHz to THz. The significant figures of merit for a structure include a high shunt impedance, a high destructive breakdown threshold, and a low fabrication cost. The beam includes the drive beam and the main beam. Without being specifically mentioned, the beam here means the electron beam. The requirements for the two beams are different because of their different roles in SWFA. The drive beam requires a higher current to establish a higher gradient, but the emphasis for the main beam is a better beam emittance.

There are two formats for the SWFA, the short-pulse Two-Beam Accelerator (TBA) and the Collinear Wakefield Accelerator (CWA). They are distinguished by whether the drive and main beams traverse the same structure. In the short-pulse TBA configuration, the drive beam energy is extracted and converted into an RF pulse in a Power Extraction and Transfer Structure (PETS), that is then used to accelerate the main beam in a separate accelerating structure. Because the drive and main beam have independent beamlines (only coupled by RF), the beam dynamics for the two beams can be treated with different strategies. More challenges are associated with the drive beamline, due to its significantly higher charge level. It is worth pointing out that the ongoing efforts toward developing a TeV-scale electron-positron linear collider from the Compact Linear Collider (CLIC) is based on the room temperature TBA scheme (http://clic.cern/, n.d.). Its choice is an RF pulse length of 240 ns and a loaded gradient of ~100 MV/m (The CLIC, 2019). Although the physics behind RF breakdown is not yet fully understood, observations in extensive experiments reveal that the RF breakdown threshold in accelerating structures increases when the RF pulse length is decreased (F. Wang, 2010) (W. Wuensch, 2003) (H. H. Braun, 2001) (Wuensch, 2002). Presently conventional high gradient rf accelerators operate with the pulse length over 100ns. For this reason, the TBA approach to an SWFA specifically means a short-pulse TBA, i.e. the RF pulse length is at the 10-ns level with >300MV/m of operational gradient.

In a CWA, the field generated by a leading high-charge drive beam (either a single bunch or a bunch train) is used to accelerate a trailing main bunch, which contains a smaller amount of charge. An important parameter that describes the performance of a wakefield accelerator is the transformer ratio R, which characterizes the efficiency of the energy transfer from the drive bunch to the witness bunch. R is less than 2 under very general conditions of linearity, a rigid relativistic longitudinally symmetric drive bunch, and identical paths through the system for both drive and accelerated beams. Another important factor to be considered in wakefield accelerators is the accelerating gradient. Structures in the CWA usually operate in the THz regime to reach the GV/m level of gradient. In principle, high R and high gradient cannot be achieved simultaneously for a CWA using a given drive charge (Zholents, 2017). Thus, increasing the drive bunch charge with a specific shaped temporal profile is the key to achieve high R and high gradient. Figure 1 illustrates the scope of SWFA research.

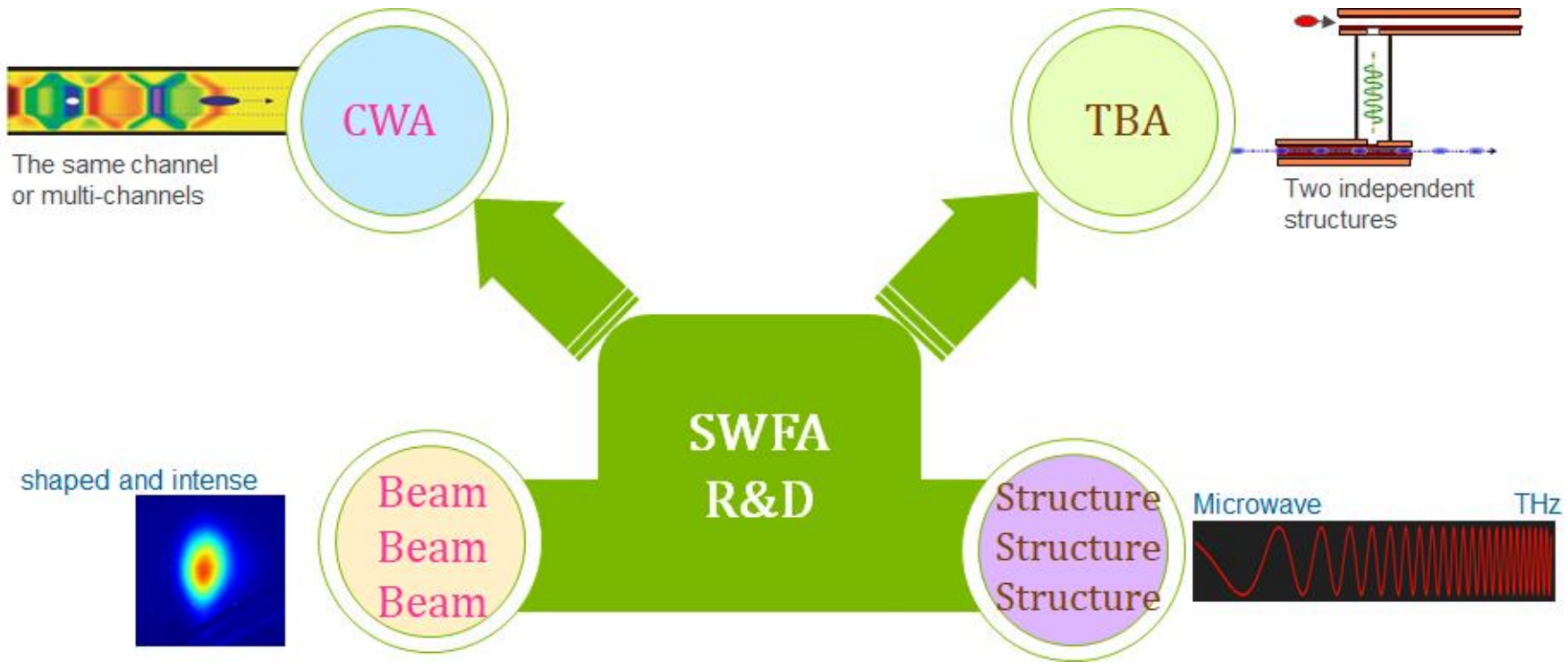

**Figure 1.** Scope of research in SWFAs.

## 1.1 Building Block of a short-pulse TBA machine

For illustrative purposes, Fig. 2 shows the core configuration of a short-pulse TBA. In a modular design of the two-beam wakefield collider (Jing, 2016), a drive beam bunch train has to be routed to its respective modules by a fast kicker; the RF power extracted from the drive beam would then be delivered to the accelerating sections for acceleration of the main beam by low-loss delay lines.

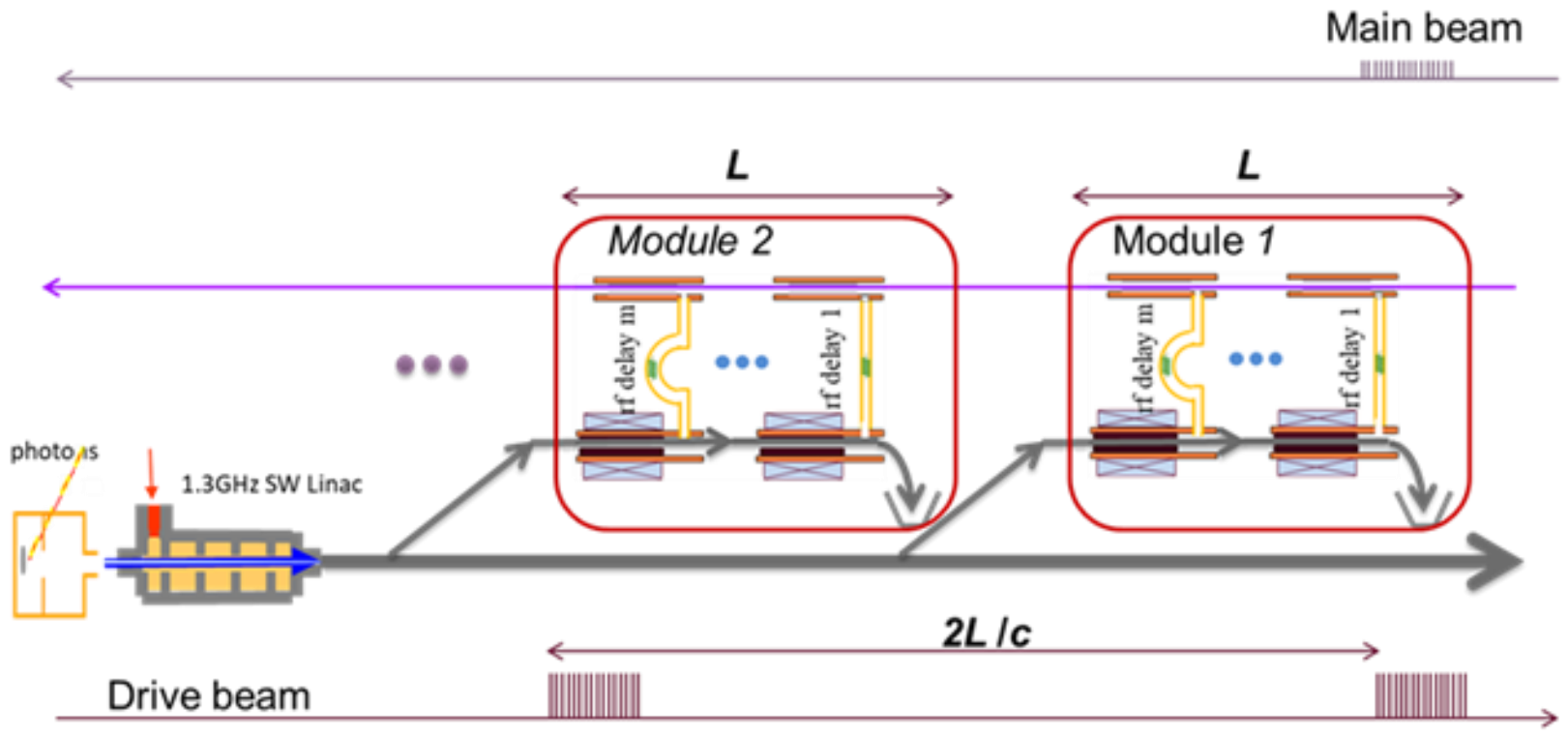

**Figure 2.** The core configuration of the short-pulse TBA, which is a building block for a linear collider.

In general, the use of staged acceleration is required by the nature of any wakefield acceleration approach to accelerate the main beam to TeV level energy. To implement staging, earlier SWFA collider concepts relied on bending the drive beams out of the drive train into power extractor units to synchronize the arrival time of the witness beam in each module (W. Gai, 2012). However, the use of a 180° arc in the drive beamline will deteriorate the drive beam quality because of coherent synchrotron radiation. The configuration shown in Fig. 2 uses RF delay lines as a substitute for the 180° bend.

## 1.2 Configuration of a Large-scale CWA Facility

In order to illustrate the key features of the CWA, **Error! Reference source not found.** shows the typical layout of a CWA beamline configuration. The scheme consists of numbers of consecutive CWA modules sharing one drive beam. The drive beam is provided by a CW SRF linac in order to achieve the high repetition rate required for a high average beam current. The drive bunch charge and the specific longitudinal profile are chosen to produce the required acceleration gradient while also achieving a high transformer ratio. The CWA can be very compact because of the THz or sub-THz operating frequency and the use of permanent quadrupole channels (A. Zholents e. a., 2016). **Error! Reference source not found.** also shows that the drive beam delay line and distribution system that are needed to enable the staged acceleration. This approach is conceptually similar to the beam plasma wakefield acceleration scheme (E. Adli, 2013).

Two major factors have to be considered in the choice of the main bunch parameters: 1) to minimize beam loading, the main beam bunch charge was made as high as possible without causing a serious drop in the net gradient; 2) to maximize beam brightness, we kept the charge as low as possible to yield small energy spread and a short bunch. A short bunch makes it easier to minimize the energy spread. The self-wake from the main beam depends on the wakefield structure, is proportional to the charge, and is inversely proportional to the bunch length (to some extent). In general, while the structure is optimized to maintain a single-mode excitation for the drive beam, multiple modes are excited by the main beam due to its shorter bunch length. But the self-wake inside the main beam will converge rapidly because the loss factor of HOMs becomes negligible after the first few higher order modes. Another critical parameter of an accelerator is the efficiency. The overall efficiency of the proposed scheme can be roughly estimated by cascading the efficiencies of each subsystem, for example, klystron AC-to-RF efficiency, RF-to-drive beam efficiency, and drive-to-main beam efficiency, etc. The drive-to-main beam efficiency needs to be close to 50% in order to have greater than 10% overall wall-plug efficiency, with consideration of the power consumption for the cooling, the beam delivery system, infrastructure, etc.

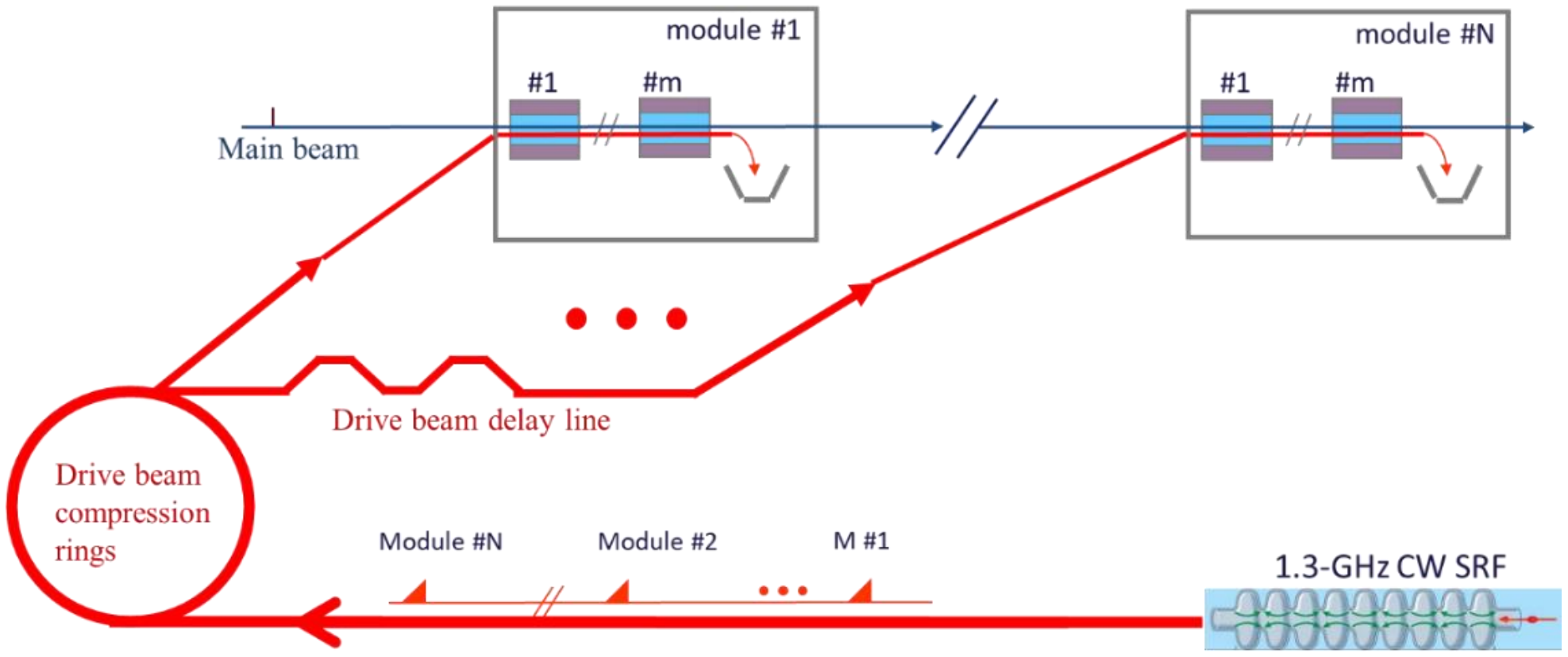

**Figure 3.** Block diagram of a 100-GeV CWA concept.

## 2. Current Status

The primary goal of SWFA research is to carry out the long-term R&D needed to realize a multi-TeV collider, and determines that the two major milestones required for structures in the short-pulse TBA approach are 300 MV/m of acceleration gradient in the accelerator and 1 GW of RF power generation in the power extractor. Figure 2 shows the SWFA-TBA progress over the past two decades. 300 MV/m of accelerating field gradient was achieved in 2020 and 500 MW of RF power generation was reached in 2021. The rapid increase of gradient and RF power are thanks to the AWA facility upgrade in 2014 to 2018, which increased the drive beam energy from 15 MeV to 65 MeV, and has been capable of producing bunch trains of 8 or 16 bunches with up to 60 nC per bunch.

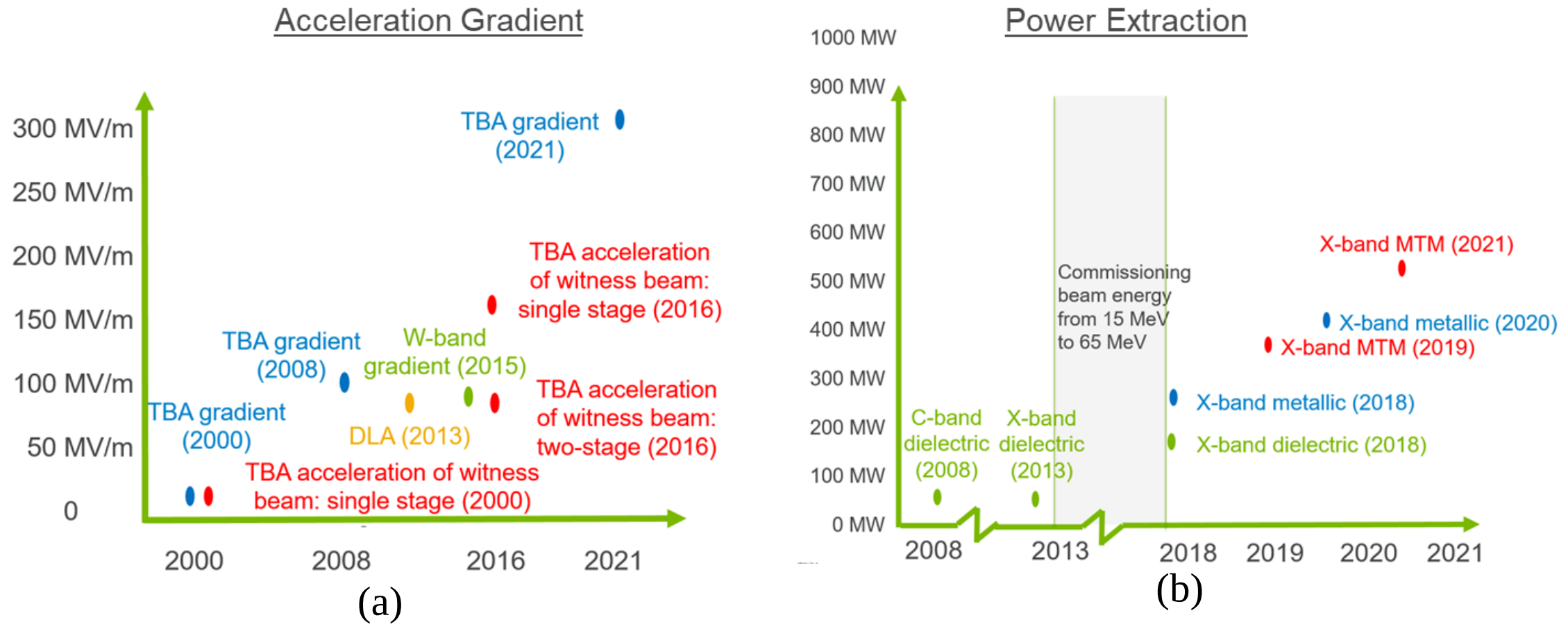

**Figure 2**. History of accelerator gradient a) and power generation b) in the short-pulse TBA scheme.

The current research in this area concentrates on exploring the latest observed Breakdown Insensitive Acceleration (BIA) regime. Over the past two decades, studies indicate that the RF breakdown process in structures operated at room temperature is predominantly correlated with the electric field, pulse width, and pulse heating (Valery Dolgashev, 2010) (Lisa Laurent, 2011) (A. Grudiev, 2009). The concept that a shorter pulse length will support a higher gradient is broadly accepted. The shortest RF pulse length that a high-energy machine can afford depends on a number of factors, such as the average beam current, efficiency, operating frequency, repetition rate, etc. However, considering just the gradient, very recent experiments carried out at the AWA facility and Tsinghua University have provided evidence that a new acceleration regime may occur when the RF pulse length is decreased to the sub-10-ns level (Power, 2021). In this new BIA regime, a breakdown event does not disrupt the accelerating field established by the incident ultra-short duration RF pulse. In other word, the short accelerating pulse is insensitive to the breakdown effect. A plausible physical explanation for the BIA regime is that the site of the breakdown and the resulting cold metallic ions migrate at a slow speed and produce the avalanche at times on the order of 10 ns after emission at which point the RF pulse for acceleration is already over. It should be pointed out that similar phenomena have been observed in different structures. A gradient of >350 MV/m was measured in both high-power RF tests and a beam generation experiment. The next step is to confirm the BIA regime acceleration for long accelerating structures. A design for fast-filling traveling-wave accelerating structures was suggested in reference (Jiang, 2021).

Up to the present time, the efforts in CWA research have been put into achieving the two main figures of merit: transformer ratio and acceleration gradient. A very high accelerating gradient (GV/m level) can be achieved in wakefield accelerators (M. C. Thompson, 2008), as well as over 300 MeV/m of beam acceleration (B. D. O'Shea, 2016). However, unless the transformer ratio R is increased far beyond the normal limit of 2, collinear wakefield acceleration suffers from low efficiency (under the assumption that the RF (or wakefield)-to-Beam efficiency is determined by the beam-loading, which a trade-off between efficiency and beam quality). The first experiment that broke the barrier of R=2 traces back to 2007, where a ramped bunch train technique was used (C. Jing e. a., 2007). Later, R=3.4 was demonstrated using the same technique in 2011 (C. Jing e. a., 2011). The significant enhancement of the transformer ratio benefitted from the steady progress of novel beam shaping techniques, which led to a record of R=5 demonstrated in an SWFA in March 2018 (Q. Gao, 2018).

One of the focuses of present research is to merge the two figures of merit in one experiment. However, these two figures of merit are not independent in a CWA. In fact, they are tied together in an unfavorable way, i.e., higher R means a lower gradient for a given drive-bunch charge (Zholents, 2017). The only solution to achieve both high gradient and high transformer ratio is to increase the charge of the drive bunch. The control of BBU by using a high-charge drive bunch was established in (Baturin & Zholents, 2018) (Li, et al., 2014). However, up to the present day, the implementation of a high charge (10-nC level) shaped bunch still remains a major obstacle. The development of a technique that is capable of shaping high-charge bunches (10-nC range, or equivalently kA range in pulsed current) is the enabling tool to reach the next major milestone.

Another research focus is the demonstration of sustainable CWA acceleration. To date, as a result of the Beam Breakup (BBU) instability in the drive beam, the ultrahigh gradient in a CWA structure has barely reached a sustainable acceleration beyond a ~10-cm structure length. The demonstration of sustainable acceleration in a meter-scale CWA structure is a high priority compared to other wakefield acceleration schemes. The use of a strong FODO lattice over the structure has been studied to control BBU. The case of a 20-meter-long, 300-GHz DWA for a MHz repetition rate X-ray FEL has been studied, including end-to-end beam simulation and a quadrupole alignment tolerance study (A. Zholents e. a., 2016). As the first step, in the near term, researchers are moving to demonstrate a half-meter-long quadrupole channel, made by stacking ten quadrupoles with alternating focusing and defocusing properties, as shown in Fig. **Figure** 5. This full-featured CWA design can be scaled up, and may open up an opportunity for the first real application of an SWFA due to its compactness, low-cost operation, and high efficiency.

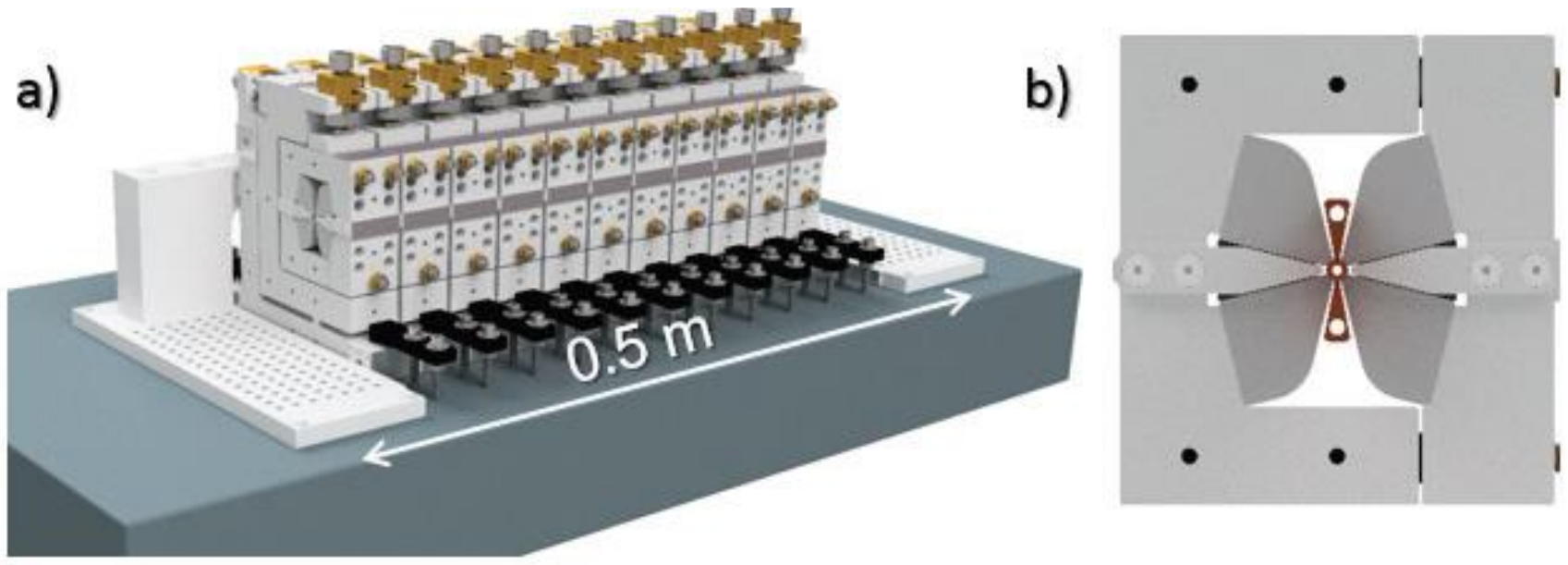

**Figure 5**. An accelerator module (a) and its front view (b). This figure is reproduced from (A. Zholents e. a., 2018).

# 3. SWFA Collider Concept

## 3.1 Argonne Flexible Linear Collider (AFLC)

We propose a 3TeV high energy linear collider based on a short rf pulse (~22ns flat top), high gradient (~267MV/m of the loaded gradient), high frequency (26GHz) short pulse two beam accelerator scheme (Figure 6). This scheme is a modular design and its unique locally repetitive drive beam structure allows a Flexible modular configuration to meet different needs. Preliminary study shows an efficiently (~14% overall efficiency) short pulse collider may be achievable. Major components are:

1. High current drive beam accelerator, including klystrons (already commerical available) and conventional standing wave linac.
2. High gradient accelerating structures to sustain 300 MV/m at 20 ns pulse length
3. High poweer extraction devices that capable of > 1 GW at 20 ns pulse length.
4. Positron production and damping rings.
5. **Final beam delivery system.**

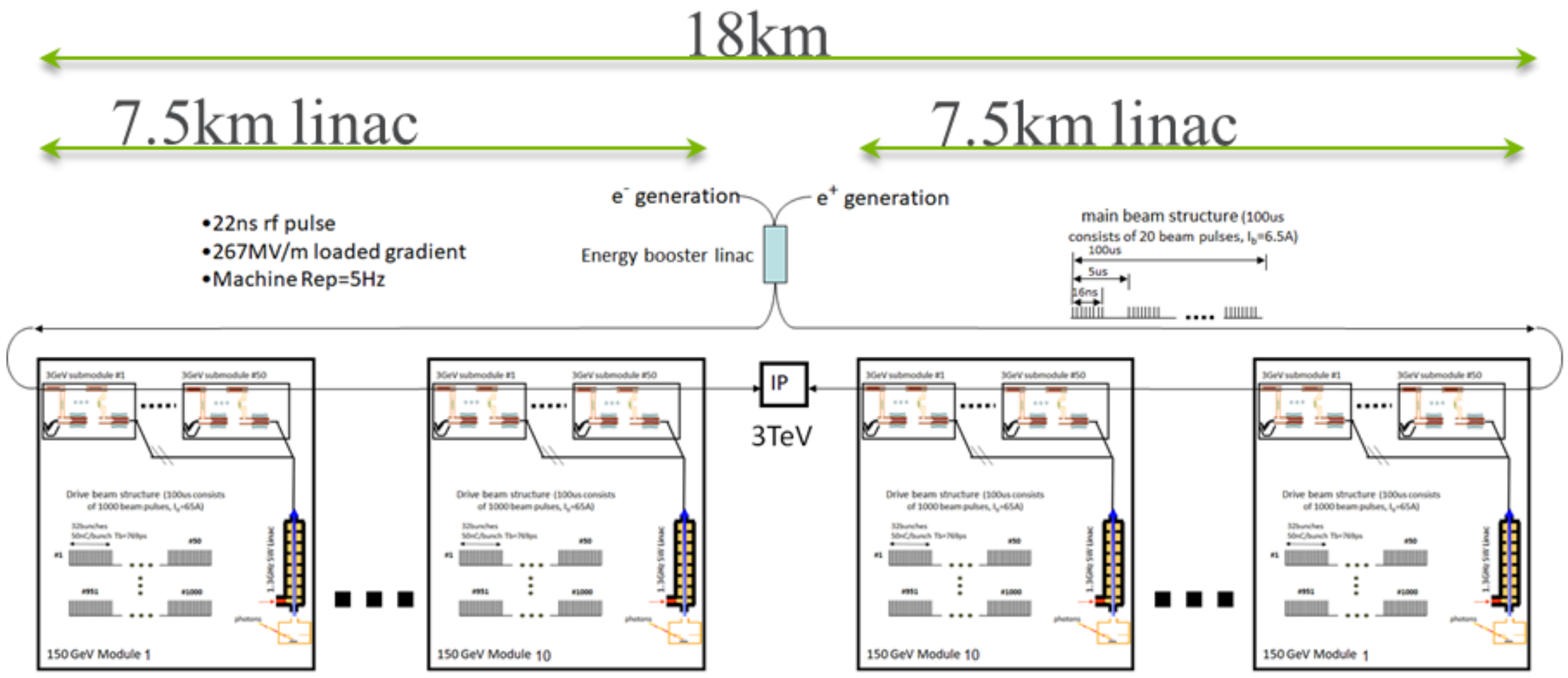

**Figure 6.** Conceptual layout of Argonne 3TeV Short Pulse TBA Linear Collider: Argonne Flexible Linear Collider (**AFLC**) . It consists of ten 150GeV stages in one side of machine. Each 150GeV stage is made up of fifty discrete 3GeV modules sharing with one drive beam source, which makes it look like the CLIC scheme except for a few critical differences. Firstly, in each 150GeV stage of the proposed short pulse TBA scheme, 1000 (=50×20) short (~24ns) micro drive pulses go through 50 modules with a local beam pulse repetition rate of 20 (it represents 20 5μs-long macro bunch train as well). Each module provides 3GeV gain, which adds up to 150GeV after 50 modules. Overall, these 1000 micro drive beam pulses, which is organized by 20 repetitive macro 5μs long macro bunches, form a 100μs giant beam pulse. On top of it, machine repetition rate is 5Hz. Secondly, to match the local beam pulse repetition rate of 20, the main beam consists 20 short beam pulses in the same 100μs period of time. Another obvious difference from the CLIC scheme is that the drive beam in this short pulse scheme is generated by a 1.3GHz rf photoinjector in each 150GeV stage with a high QE cathode, which can provide 50nC/bunch with a bunch separation of 769ps (32 sequential bunches form a ~24ns micro drive pulse). At last, to achieve a high rf-to-beam efficiency in the main linacs under the short rf pulse condition, we choose a high frequency (26GHz), high group velocity (~11%c), and broadband structure, which in turns provide ~270MV/m of the loaded gradient, ~9ns filling time, and ~3ns rise time.

## Major Parameters:

**Table 1**. Some preliminary design parameters of 3-TeV short pulse linear collider

| | |
|---|---|
| Main linac frequency | 26GHz |
| Drive linac frequency | 1.3GHz |
| Main linac loaded gradient | 267MV/m |
| Main beam current (in pulse) | 6.5A |
| Machine repetition rate | 5Hz |
| Average drive beam current (one side) | 80mA |
| Average drive beam power (one side) | 68.8MW |
| Average main beam current (one side) | 9.27μA |
| Average main beam power (one side) | 13.9MW |

**Table 2**. Parameters of 26GHz Dielectric Based Wakefield Power Extractor.

| **Geometric and accelerating parameters** | **value** |
|---|---|
| ID / OD of dielectric tube | 7 mm /9.068 mm |
| Dielectric constant | 6.64 |
| Length of dielectric tubes | 300 mm |
| Vg | 0.254c |
| R/Q | 9.8 kΩ/m |
| Rf pulse rise time | 2.9 ns |
| $BW_{_3dB}$ of the requested coupler | 120MHz |
| Steady power (50nC/bunch, $\sigma$ z=1mm) | 1.33 GW |
| RF pulse duration (32 bunches) | 22ns (flat top) |
| Peak Gradient | 167MV/m |
| Energy loss of the beam in the steady state | 20.5MeV |

**Table 3**. Parameters of 26GHz Dielectric Based Accelerator.

| **Geometric and accelerating parameters** | **value** |
|---|---|
| ID / OD of dielectric tube | 3 mm /5.025 mm |
| Dielectric constant | 9.7 |
| Length of dielectric tubes | 300 mm |
| Vg | 11.13%c |
| Tfill | 9ns |
| R/Q | 22 kΩ/m |
| Q (loss tan=10^-4) | 2295 |
| Shunt impedance | 50.4 MΩ/m |
| $BW_{_3dB}$ of the requested coupler | 120 MHz |
| Eacc for 1.26GW input | 316 MV/m |
| Eload for 1.26GW input | 267 MV/m |

**AFLC Approaches to Fundamental Feasibility Issues of LC:**

1. Positron Production and Acceleration---- AFLC adapts the positron production and damping ring similar to CLIC design. Over the past several years, driven by the needs of the international linear collider design, AWA have provided leadership in technical area of polarized positron source studies for ILC Global Design Effort (GDE). AWA has performed a comprehensive study of both the undulator-based and the Compton-based e+ sources. We have made design and prototyping of major ILC e+ components for a 300 GeV center of mass machine. The experience can be extended naturally to TeV level.

2. Acceleration Scalability (Staging)---- AFLC adapts the novel staging scheme based on mature RF delay lines techniques. Staging is one of the most important issues for all wakefield acceleration concepts. In general design, a series of 180 degree bending arcs in the drive beamline is needed to continuously accelerate the witness bunch to an ultrahigh energy in relays. The large bending arcs or magnetic chicane type of drive beam delays suffer from quality degradation of the high charge drive beam due to Coherent Synchrotron Radiation (CSR). AFLC uses different rf delays among the TBA pairs to synchronize the timing between drive and main beam so that the need for the drive beam bending is avoided (refer to Fig.2 for details).

3. Wall Plug Efficiency----- AFLC has an estimated wall plug efficiency of 14%. Major sub-efficiencies in the power flow chain to calculate the overall efficiency include efficiency from the wall plug to power supplies to klystrons rf output (80%, considering high efficiency klystrons), rf-to-drive beam efficiency (86%), efficiency of the wakefield power extraction (96%), and rf-to-main beam efficiency (58%, considering main beam bunch shaping and high shunt impedance structures). Other power consumption including the main beam injection, magnets, services, infrastructure, and detector are also accounted in. The total delivered beam power is 27.8MW. The total wall plug power is estimated 195MW.

4. Sustainability of Acceleration (beam instability) -----AFLC is a structure-based scheme. All decelerators and accelerators are using transverse damping slots to minimize the risk of BBU (beam breakup). In particular, it has been demonstrated that the transverse mode damping in dielectric based accelerating structures can be implemented in a lower cost without losing its effectiveness.

5. Fabrication Cost--------AFLC is presumably designed using dielectric accelerating structures in the current stage. The fabrication cost of dielectric structures is significantly lower due to the nature of dielectric-lined waveguide as a slow wave structure.

## 3.2 The CWA steppingstone facility

*The CWA driver application* is a strawman design of a high repetition rate multi-user X-ray FEL (Fig.7) and is developed in collaboration with the Argonne APS (A. Zholents e. a., 2018) (A. Zholents e. a., 2016). The AWA program is focused on the preliminary design of the multi-meter long collinear wakefield accelerator to achieve a highly efficient transfer of the drive bunch energy to the wakefields and to the witness bunch is considered. The CWA linac is made from ~0.5 m long accelerator modules containing a vacuum chamber with corrugated waveguides as the walls, a quadrupole wiggler, an rf coupler, and BPM assembly. The single bunch breakup instability is a major limiting factor for accelerator efficiency, and the BNS damping in the strong wakefield regime is applied to obtain the stable multi-meter long propagation of a drive bunch. This CWA application is considered as a steppingstone facility towards the exploration of a CWA-based linear collider.

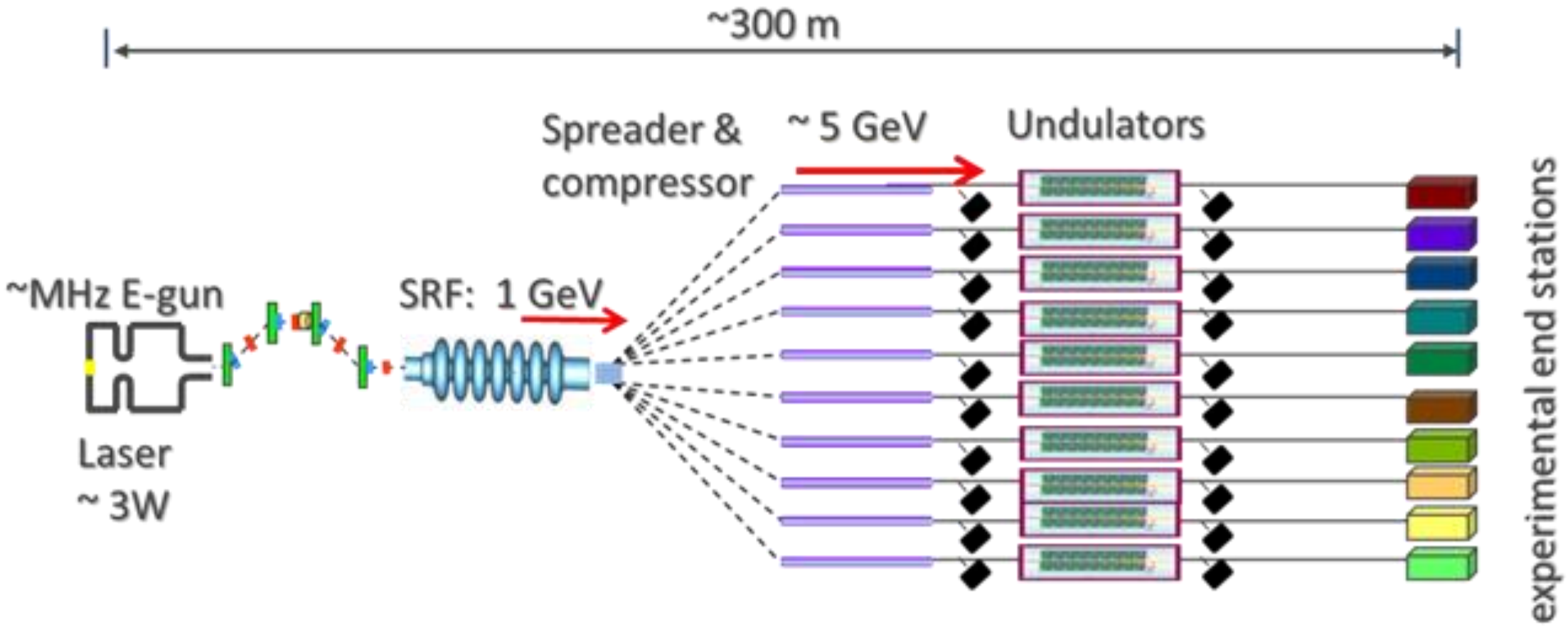

**Figure 7**. The CWA scheme of a high repetition rate X-ray FEL facility for the SWFA technology.

# 4. Development Path

## 4.1 Roadmap of Short-Pulse TBA R&D

The first roadmap for the SWFA concept was formulated in 2016 (Advanced Accelerator Concepts Research Roadmap Workshop Report, 2016), and in the past five years, steady progress has been made. Based on new results and new technologies as well as newly allocated resources, by the time of the Snowmass 2022 meeting, we may lay out a strategic plan of SWFA for the US high-energy physics research in the coming decade, and will suggest an updated 10-year roadmap for the short-pulse TBA approach to a TeV-class linear collider. Figure 8 is by no means of a formal SWFA roadmap, but it may shed light on this research area from the authors' perspective. In this roadmap, the Technical Consolidation Phase aims to mature the critical technical elements, including GW power production, 300 MV/m gradient acceleration, staging, the main beam source, etc. Then, it migrates into the Technical Integration Phase, where we do prepare for a 10-GeV-level facility. It may be an X-FEL facility or a similar kind.

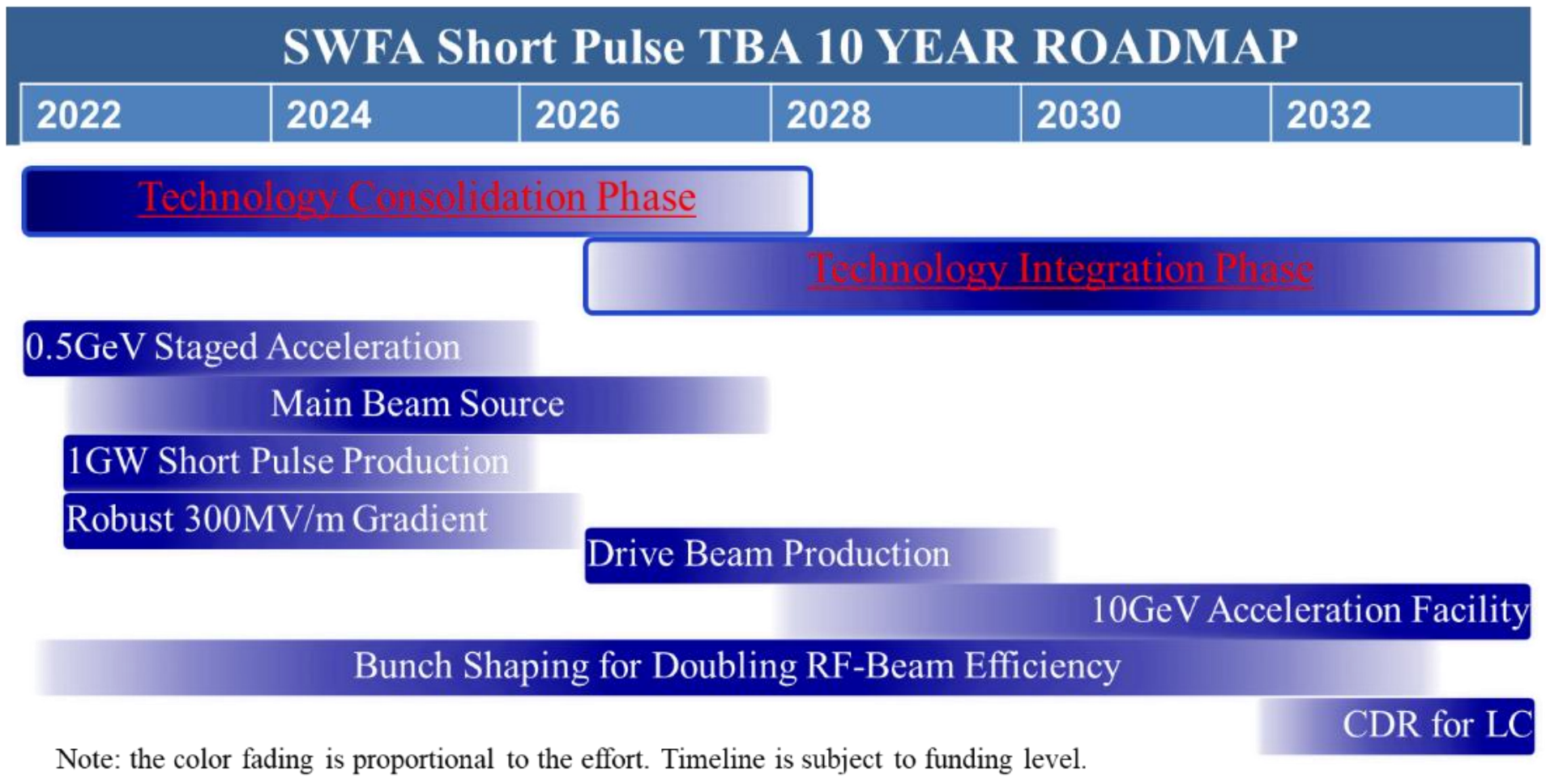

**Figure 8**. The roadmap for the SWFA short-pulse TBA concept suggested by the authors. The color fading is proportional to the efforts.

## 4.2 Roadmap of CWA R&D

Similar to Section 4.1, we suggest a 10-year roadmap for the CWA (See Fig.9). The goal is to develop within the first 5 years a scaled metal CWA unit that can demonstrate a significant energy gain (e.g. double 100 MeV or higher) as well preserve the quality of the main beam. Then, the roadmap moves to the demonstration of a staged CWA module. This module is the key component of the CWA for the target multi-GeV level XFEL application or for a linear collider.

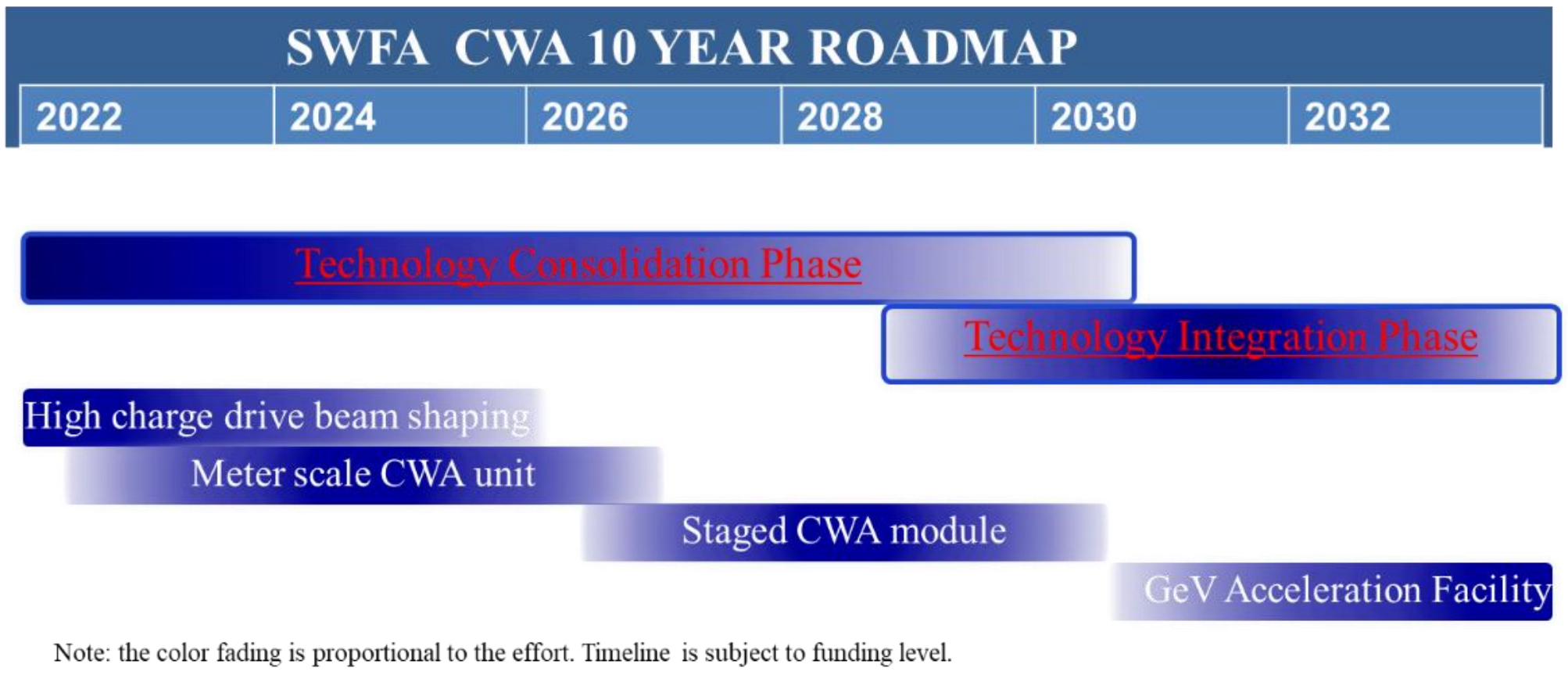

**Figure 9.** The roadmap for the SWFA CWA concept suggested by the authors.

## 4.3 Near to Mid Term Demonstrators

### 4.3.1 Near Term Demonstrators at Argonne Wakefield Accelerator facility

The most exciting work we have planned for the **SWFA 3~5year vision** is the integration of the various component technologies (for CWA and TBA) into three demonstration modules. These modules will focus the AWA research to show that all the various SWFA components both work separately and can be brought together to demonstrate a fully functioning acceleration module. Two modules will demonstrate high energy gain (a 0.5GeV demonstrator based on the TBA scheme and an energy doubler based on the CWA scheme) and one module to demonstrate high rf-to-main beam efficiency, 27% → 58%, (using the DDA structure and main bunch shaping). In summary, the cost-effective AWA program combines the unique capabilities of the AWA facility (high charge, beam shaping, and independent and easily reconfigurable beamlines), strong group expertise, need to advance the AAC roadmap.

Major Deliverables:

I. **TBA: 500 MeV demonstrator (Fig.10).** This module can be installed in the existing AWA facility bunker with no need for expansion. It is the upper limit of what can be accomplish inside the current facility and its successful demonstration would represent a major milestone in AAC roadmap.  It will use four pairs of 26 GHz dielectric-disk structures installed into two separate stages.  The four power extraction and transfer structures (PETS1-4) will each generate 1.2 GW and the four accelerating structures (ACC1-4) will operate at an acceleration gradient of 250 MeV/m.  All structures will be based on new idea of dielectric disks embedded in a metallic waveguide.

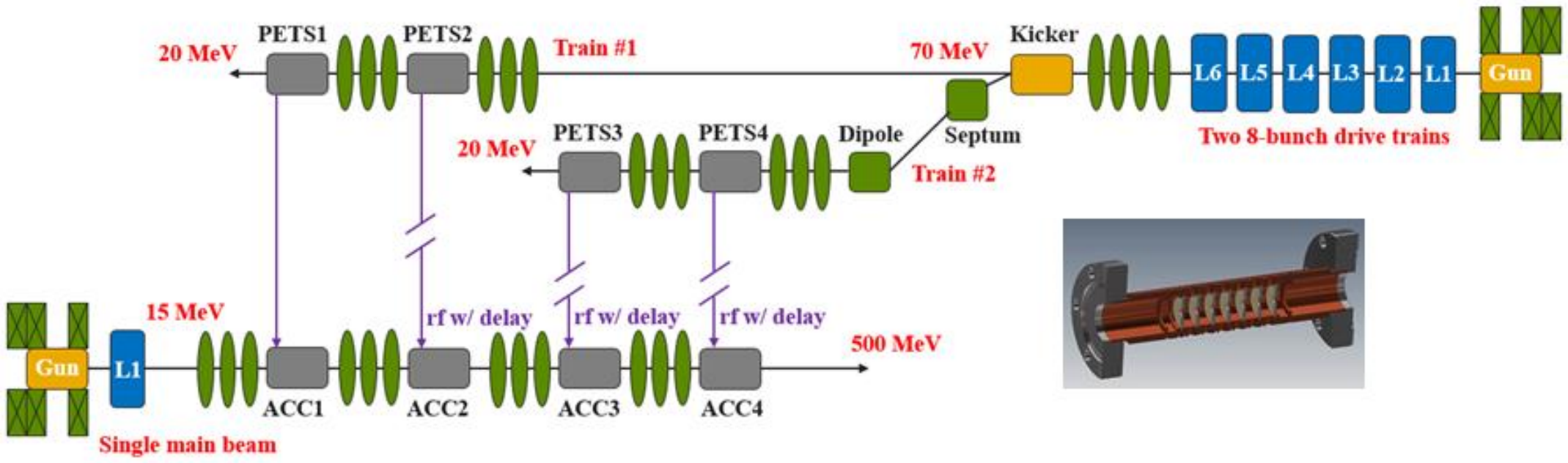

**Figure 10**. Layout of the proposed 0.5GeV demonstrator module in the AWA facility. The inset figure is a cut-away view of dielectric disk accelerator, one type of dielectric accelerators with high shunt impedance and high group velocity.

II. **TBA: Efficiency doubler. (Fig. 11**). This module can be installed in the existing AWA facility bunker with no need for expansion.   Its successful demonstration would represent a major milestone in AAC roadmap.  It will require the development of both the structure and beam technology to demonstrate the doubling of the efficiency (from 27% to 58%).   Since the main beam needs to be shaped, we will first consider using the "drive" linac to send a beam into the EEX beamline to generate the shaped main beam and use the "witness" linac to generate the drive beam.

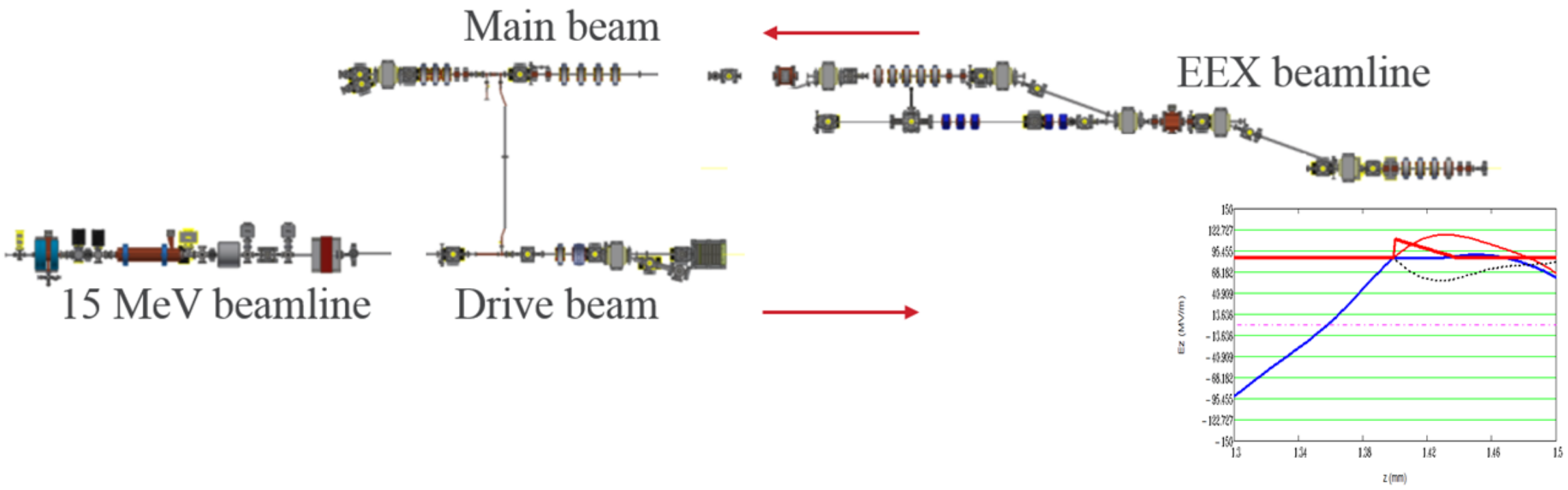

**Figure 11**. Layout of the proposed efficiency doubler. The inset figure shows the principle of a shaped main beam for increasing beamloading without sacrificing the beam energy spread.

III. **CWA: energy doubler (Fig.12).** This module can be installed in the existing AWA facility bunker with no need for expansion. The main deliverable for this module is to double the AWA witness bunch energy in 1 m, where the witness bunch be generated with the drive linac and therefore have an initial energy of approximately 50-70 MeV depending on the linac optimization. Therefore, the goal would be to take the witness bunch energy from ~50 MeV and increase by a factor of 2-3 times, to an energy of 100-150 MeV. This module is the key component of the CWA for the target XFEL application.

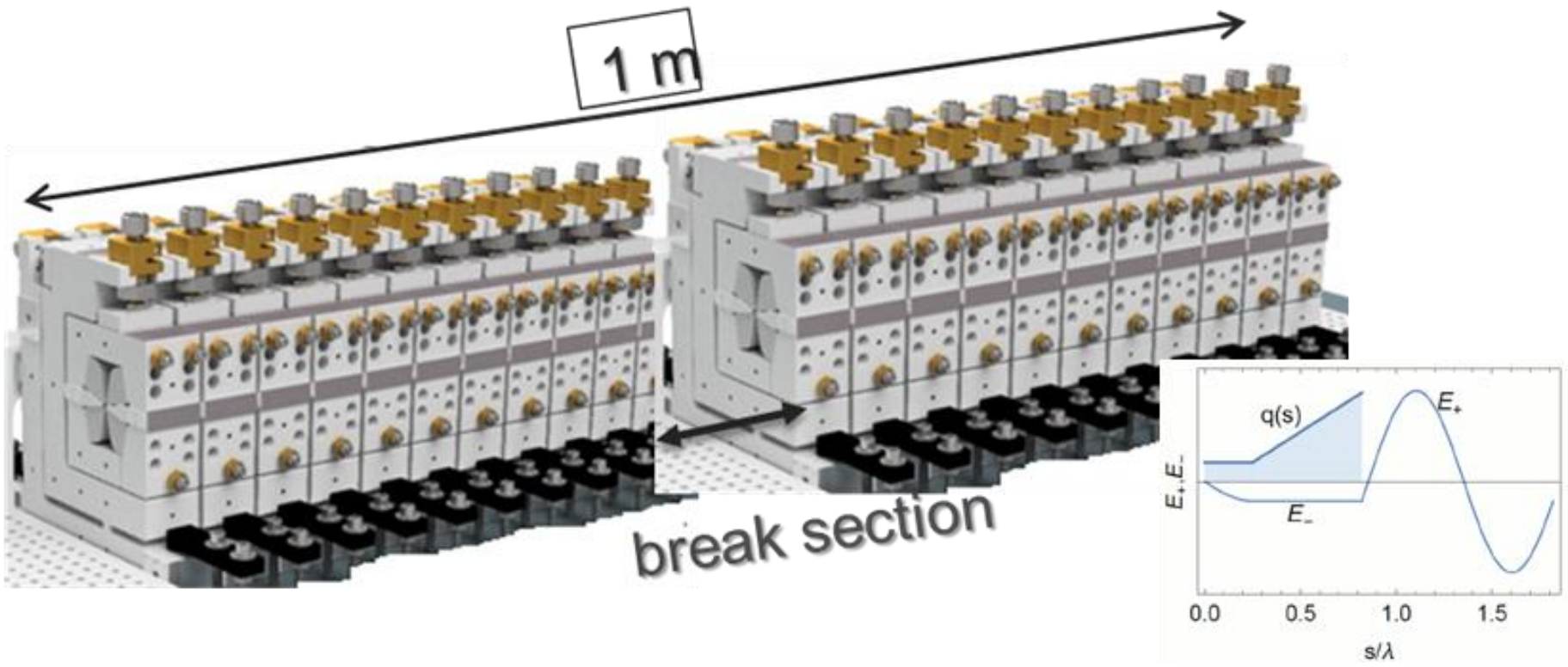

**Figure 12**. The engergy doubler demonstrator. The figure is reproduced from (A. Zholents e. a., 2018). The inset figure represents the 10nC door step beam that is needed for this demonstrator.

### 4.3.2 Mid-Term Demonstrators

After the near-term demonstrators are successful, we will seek a major upgrade in the **next 10 years** to boost the drive beam and main beam energy towards a fully functional GeV-scale module. The primary target is a 3GeV module, which is an identical sub-module for future TeV-scale linear collider and XFEL.

# 5. Synergies with Others

## 5.1 Synergies to AAC community

The SWFA research has strong synergies with PWFA, some relevance to LPA, and therefore contributes to their roadmaps as well as a small effort in stewardship applications. The drive beam technology developed for CWA, whether for SWFA or PWFA, requires a single high-charge (10-100nC) e- drive bunch with a carefully shaped longitudinal profile (approximately triangular current distribution) to accelerate a single e- or e+ main bunch with both high-gradient and high transformer ratio. All CWA schemes use high transformer ratio to enable a low energy drive beam to accelerate the main beam to high energy, but as shown in Reference (Zholents, 2017), the accelerating gradient drops off very quickly as transformer ratio increases. Therefore, since the CWA will not use longitudinally short Gaussian bunches very high charge is needed to operate at high gradient even for ~THz scale SWFA or PWFA accelerators. In fact, a record high transformer ratio in PWFA has been demonstrated in the AWA beamline recently (R. Roussel, 2020), a primary SWFA testing facility. In order to operate a high-efficiency all three technologies (SWFA, PWFA, LPA) benefit from using a specially shaped main beam to optimally load the wakefield. The AWA shaping capability is ideally suited to develop and test the efficiency enhancement due to main bunch shaping.

## 5.2 Synergies to Normal conducting RF accelerators

The high charge drive used in SWFA can be used to generate high power rf beyond X-band, thus provide a platform to study the frequency scaling law of the acceleration gradient. Also, in combination of the independent witness beamline, physics and engineering issues (wakefield, beam instabilities, rf breakdown, and rf transportation, etc.) of the high frequency acceleration (W-band and beyond) can be explored.

Mutually the advancement of NCRF technologies offers great synergistic opportunities to the SWFA roadmap, in particular, the high efficiency Klystron and high power rf components.

# 6. Conclusion

In a general sense, high gradient is desirable for a TeV class linear collider design because it can reduce the length of total linacs, thus the construction cost. More importantly, the efficiency and the cost to sustain such a gradient remain in high priority as well in the optimization process of an overall design. The goal of the HEP AAC program is to develop transformational technologies to enable building a low-cost, high-efficiency machine serving HEP frontiers. In 2009, a concept of Argonne Flexible Linear Collider (AFLC) was first proposed at High Energy Physics Division of Argonne National Laboratory. The most distinguishable features of this conceptual design are its modular configuration and short rf pulse (~20ns) high gradient (~300MeV/m) operation, which make it scalable in high-fidelity from a GeV level scale stepping-stone facility to TeV class machine, and also allows a flexible configuration to meet different needs. Preliminary study shows that ~15% of efficiency may be achievable, which opens an opportunity to lower the site power below 200MW for a 3TeV collider. Since its inception, main challenges involved with AFLC have been progressively addressed in AWA facility as well as other collaborative institutes, which include efforts to develop wakefield based 1 GW pulsed power source, low cost high gradient accelerators, and staging etc. Entering the new era of Snowmass 2022, strong technological and resources are preferably requested to support research activities in all facets of AFLC moving this conceptual design to the real feasibility study. We also advocate to update the SWFA roadmap to encompass the latest development in the field, including THz acceleration in picosecond pulse regime; novel structures like metamaterial, crystal, and nanostructures; revolutionary structure fabrication techniques like 3D-printing and brazeless structures; and near-term applications, like compact X-band other applications ray sources, and other applications.